# Source normalized indicators of citation impact: An overview of different approaches and an empirical comparison

Ludo Waltman and Nees Jan van Eck

Centre for Science and Technology Studies, Leiden University, The Netherlands
{waltmanlr, ecknjpvan}@cwts.leidenuniv.nl

Different scientific fields have different citation practices. Citation-based bibliometric indicators need to normalize for such differences between fields in order to allow for meaningful between-field comparisons of citation impact. Traditionally, normalization for field differences has usually been done based on a field classification system. In this approach, each publication belongs to one or more fields and the citation impact of a publication is calculated relative to the other publications in the same field. Recently, the idea of source normalization was introduced, which offers an alternative approach to normalize for field differences. In this approach, normalization is done by looking at the referencing behavior of citing publications or citing journals.
In this paper, we provide an overview of a number of source normalization approaches and we empirically compare these approaches with a traditional normalization approach based on a field classification system. We also pay attention to the issue of the selection of the journals to be included in a normalization for field differences. Our analysis indicates a number of problems of the traditional classification-system-based normalization approach, suggesting that source normalization approaches may yield more accurate results.

## 1. Introduction

The use of citation-based bibliometric indicators for assessing the impact of scientific publications has become more and more popular. One of the most important difficulties in the development of these indicators concerns the comparison of the citation impact of publications from different scientific fields. It is well known that different fields may have very different citation practices. In fields with a high citation density (e.g., cell biology), the average number of citations received per publication may for instance be more than an order of magnitude larger than in fields with a low citation density (e.g., mathematics). Given these large differences in citation practices, the development of bibliometric indicators that allow for meaningful between-field comparisons is clearly a critical issue.

Traditionally, bibliometric indicators have usually relied on a field classification system to normalize for field differences (e.g., Braun & Glänzel, 1990; Glänzel, Thijs, Schubert, & Debackere, 2009; Moed, De Bruin, & Van Leeuwen, 1995; Waltman, Van Eck, Van Leeuwen, Visser, & Van Raan, 2011). A field classification system assigns each publication to one or more fields (e.g., biochemistry, economics, mathematics, neurology, etc.). Normalization for field differences is done by calculating the citation impact of a publication relative to all publications in the same field. The most commonly used field classification system is the system of journal subject categories in the Web of Science (WoS) database of Thomson Reuters. In this system, each journal is assigned to one or more fields. A publication belongs to the fields of the journal in which it has appeared. There are about 250 fields in the WoS subject categories system (including arts and humanities fields). Journals such as *Nature*, *PNAS*, and *Science* belong to a special 'Multidisciplinary Sciences' category.

Normalization based on a field classification system has a number of limitations. First, the idea of science being subdivided into a number of clearly delineated fields is artificial. In reality, boundaries between fields may be rather fuzzy. Second, fields can be defined at different levels of detail, and given a certain level at which one has defined one's fields, it is always possible to go one level deeper and to define subfields at this deeper level. It is quite well possible that the subfields within a single field differ significantly from each other in terms of citation practices (e.g., Adams, Gurney, & Jackson, 2008; Neuhaus & Daniel, 2009; Van Leeuwen & Calero Medina, 2012; Waltman, Yan, & Van Eck, 2011; Zitt, Ramanana-Rahary, & Bassecoulard, 2005). Hence, in many cases, it is not clear to what extent fields can be regarded as homogeneous entities. Third, in the case of a field classification system defined at the level of journals rather than individual publications, there is the problem of journals with a broad scope, not only journals such as *Nature*, *PNAS*, and *Science*, but also for instance *Journal of the American Chemical Society*, *New England Journal of Medicine*, and *Physical Review Letters*. These journals do not fit neatly into a field classification system.

Recently, an alternative approach to normalization for field differences was introduced in the literature. This approach is referred to as citing-side normalization (Zitt, 2010, 2011; Zitt & Small, 2008), source normalization (Moed, 2010; Waltman & Van Eck, 2010a), fractional counting of citations (Leydesdorff & Bornmann, 2011; Leydesdorff & Opthof, 2010; Leydesdorff, Zhou, & Bornmann, in press; Zhou & Leydesdorff, 2011), or a priori normalization (Glänzel, Schubert, Thijs, & Debackere, 2011). In this paper, we use the term 'source normalization'. The source normalization approach does not require a field classification system. Instead, it starts from the idea that the main reason for differences in citation density between fields is that in some fields publications tend to have longer reference lists than in others. In fields with long reference lists, it can be expected that on average publications are cited more frequently than in fields with short reference lists. Based on this idea, the source normalization approach aims to normalize for field differences by correcting for the reference list length of citing publications or citing journals.

In this paper, we discuss and compare a number of approaches that can be taken to normalize for field differences. Our focus is on source normalization approaches. We include three source normalization approaches in our analysis, one based on the idea of the audience factor of Zitt and Small (2008), one based on the idea of fractional citation counting introduced by Leydesdorff and Opthof (2010), and one based on the idea of the revised SNIP indicator of Waltman, Van Eck, Van Leeuwen, and Visser (2012). The three source normalization approaches are compared empirically with a traditional normalization approach based on a field classification system. In addition to the issue of the choice of a normalization approach, we also consider another, often overlooked issue, namely the issue of the selection of the journals (or the publications) to be included in a normalization for field differences. For instance, should a normalization be based simply on all journals available in a bibliographic database (including trade journals, popular magazines, scientific journals with a strong national orientation, etc.) or should it be based on a selection of journals, such as all international scientific journals? The empirical analysis that we present uses the various normalization approaches to assess the citation impact of journals in the WoS database. Although we focus on assessing the impact of journals, we emphasize that the normalization approaches we study can also be used for assessing the impact of universities, research groups, individual researchers, etc.

The organization of this paper is as follows. Section 2 discusses the issue of the selection of the journals to be included in a normalization for field differences. Section 3 introduces the bibliometric indicators that we study and the corresponding normalization approaches. Section 4 presents the results of our empirical analysis. Finally, Section 5 summarizes our conclusions.

## 2. Selection of journals

Usually, in a normalization for field differences, all journals available in a bibliographic database such as WoS or Scopus are included. However, in addition to 'regular' scientific journals that aim to serve an international community of researchers, these databases also cover a significant number of 'special' journals, often with a low citation impact. Examples include trade journals targeted primarily at an industrial rather than a scientific audience and popular magazines aimed at a broad, non-expert readership.[1] Another example are scientific journals with a strong focus on a scientific community in one particular country or group of countries.[2] In many cases, including these special journals in a normalization is problematic. This is illustrated by the following example.

Suppose that a traditional normalization approach based on a field classification system is used, and consider two fields, field X and field Y. In field X, our database covers only 'regular' scientific journals. In field Y, on the other hand, our database also covers a number of 'special' journals, for instance trade journals and national scientific journals. Suppose that, compared with the regular journals in field Y, the special journals in this field receive very few citations. It may now be argued that in the normalization for field differences the regular journals in field Y have an advantage over the regular journals in field X. This is because in the normalization the citation impact of a journal is compared with the citation impact of all journals in the same field. Because of the presence of a number of special journals with a low citation impact in field Y, it is relatively easy for the regular journals in this field to perform well in this comparison. This is not the case for the regular journals in field X, and these journals may therefore be argued to have a disadvantage compared with their counterparts in field Y. To get rid of this disadvantage, the special journals in field Y would need to be excluded from the normalization.

Of course, it is rather difficult to distinguish in an accurate way between what should count as a 'regular' scientific journal and what should count as a 'special' journal. In this paper, it is not our aim to introduce precise criteria for making this distinction. However, we do want to explore the consequences of excluding certain types of journals from a normalization. Our focus is on excluding journals that are strongly oriented on one or a few countries (see also Zitt, Ramanana-Rahary, & Bassecoulard, 2003). We refer to these journals as national and regional journals.

---

[1] WoS covers a substantial number of trade magazines. Examples of some of the larger ones are *Genetic Engineering & Biotechnology News*, *Naval Architect*, and *Professional Engineering*. Popular magazines covered by WoS include, among others, the scientific magazines *American Scientist*, *New Scientist*, and *Scientific American* and the business magazines *Forbes* and *Fortune*.

[2] In the case of the Netherlands, WoS for instance covers the Dutch language journals *Psychologie & Gezondheid*, *Tijdschrift voor Communicatiewetenschap*, and *Tijdschrift voor Diergeneeskunde* as well as the English language journals *Economist-Netherlands*, *Netherlands Heart Journal*, and *Netherlands Journal of Medicine*.

How can national and regional journals be distinguished from international journals? In this paper, we try to distinguish between these two types of journals by analyzing the countries mentioned in the address lists of the publications of a journal. More specifically, for each combination of a journal and a country, we count the number of times the country is mentioned in the address lists of the journal's publications. In this way, we obtain for each journal a distribution over countries. If for a given journal this distribution is strongly concentrated on one or a few countries, this is a clear indication that the journal has a national or regional orientation. Mathematically, to determine the degree to which a journal has a national or regional orientation, we compare the journal's distribution over countries with the overall distribution obtained based on all journals in the database (see also Zitt & Bassecoulard, 1998). We use the Kullback-Leibler divergence for comparing the two distributions. For a given journal $i$, the Kullback-Leibler divergence equals

$$d_i = \sum_j p_{ij} \ln \frac{p_{ij}}{q_j}, \qquad (1)$$

where $p_{ij}$ denotes the proportion of the addresses in journal $i$ that are from country $j$ and $q_j$ denotes the proportion of all addresses in the database that are from this country. The higher the value of $d_i$, the stronger the national or regional orientation of journal $i$. Some threshold for $d_i$ needs to be chosen to determine the boundary between what counts as a national or regional journal and what does not.

## 3. Indicators

Five bibliometric indicators are considered in our analysis. One indicator does not normalize for field differences, one indicator uses a traditional normalization approach based on a field classification system, and the other three indicators each use a different source normalization approach. The indicators are used to assess the citation impact of journals in the WoS database. The period of analysis has a length of four years. All citations received during the four-year period by publications that appeared in the first three years are counted. This means that publications from the first year have a four-year citation window, while publications from the second and the third year have, respectively, a three-year and a two-year citation window. No citations are counted for publications from the fourth year. The four normalized indicators aim to normalize not only for field differences but also differences in citation window length.

The *mean citation score* (MCS) indicator is the simplest indicator in our analysis. The indicator does not normalize for field differences or differences in citation window length and simply equals the average number of citations a journal has received per publication. The MCS value of a journal can be written as

$$\text{MCS} = \frac{n}{m}, \qquad (2)$$

where $n$ denotes the total number of citations received by the journal and $m$ denotes the number of publications of the journal. The MCS indicator is similar to the journal impact factor, but unlike the journal impact factor the MCS indicator uses multiple citing years.

The *mean normalized citation score* (MNCS) indicator normalizes for field differences and differences in citation window length (Waltman, Van Eck et al., 2011; see also Lundberg, 2007). The normalization for field differences is based on a field classification system. In our analysis, the system of WoS subject categories is used. The MNCS value of a journal is calculated as

$$\text{MNCS} = \frac{1}{m}\left(\frac{n_1}{e_1} + \frac{n_2}{e_2} + \ldots + \frac{n_m}{e_m}\right), \qquad (3)$$

where $n_i$ denotes the number of citations of the $i$th publication of the journal and $e_i$ denotes the average number of citations of all publications in the journal's field in the year in which the $i$th publication appeared.[3] Interpreting $e_i$ as the 'expected' number of citations of the $i$th publication, $n_i / e_i$ denotes the ratio of the $i$th publication's actual and expected number of citations. A ratio above (below) one indicates that the number of citations of the publication is above (below) what would be expected based on the field and the year in which the publication appeared. The MNCS value of a journal equals the average of the actual/expected ratios of the journal's publications. An MNCS value above (below) one means that on average the publications of the journal are cited more (less) frequently than would be expected based on their field and publication year.

We now turn to the three indicators that use a source normalization approach. We refer to these indicators as MSNCS[(1)], MSNCS[(2)], and MSNCS[(3)], where MSNCS stands for *mean source normalized citation score*. The general idea of the three MSNCS indicators is to calculate a journal's average number of citations per publication, where each citation is weighted based on the referencing behavior of the citing publication or the citing journal. The three MSNCS indicators differ from each other in the exact way in which the weight of a citation is determined. An important concept in the case of all three indicators is the notion of an active reference (Zitt & Small, 2008). An active reference is a reference that falls within a certain reference window and that points to a publication in a journal covered by one's database. For instance, in the case of a four-year reference window, the number of active references in a publication from 2008 equals the number of references in this publication that point to publications from the period 2005–2008 in journals covered by the database.

The MSNCS[(1)] value of a journal is given by

$$\text{MSNCS}^{(1)} = \frac{1}{m}\left(\frac{1}{a_1} + \frac{1}{a_2} + \ldots + \frac{1}{a_n}\right), \qquad (4)$$

where $a_j$ denotes the average number of active references in all publications that appeared in the same journal and in the same year as the publication from which the $j$th citation originates. The length of the reference window within which active references are counted equals the length of the citation window of the publication by which the $j$th citation is received. The following example illustrates the definition of $a_j$. Suppose that the period of analysis is 2008–2011, and suppose that the $j$th citation originates from a citing publication from 2010 and is received by a cited publication from 2009. Since the cited publication has a three-year citation window (i.e., 2009–2011), $a_j$ equals the average number of active references in all publications that appeared in the citing journal in 2010, where active references are counted within a three-year reference window (i.e., 2008–2010). The MSNCS$^{(1)}$ indicator is based on the idea of the audience factor of Zitt and Small (2008). However, unlike the audience factor, the MSNCS$^{(1)}$ indicator uses multiple citing years.

---

[3] In the case of a journal that is assigned to multiple fields in a field classification system, $e_i$ is calculated as the harmonic average of the expected numbers of citations obtained for the different fields. For a justification of this approach, we refer to Waltman, Van Eck et al. (2011).

The MSNCS$^{(2)}$ indicator is similar to the MSNCS$^{(1)}$ indicator, but instead of the average number of active references in the citing journal it looks at the number of active references in the citing publication. In mathematical terms,

$$\text{MSNCS}^{(2)} = \frac{1}{m}\left(\frac{1}{r_1} + \frac{1}{r_2} + \ldots + \frac{1}{r_n}\right), \tag{5}$$

where $r_j$ denotes the number of active references in the publication from which the $j$th citation originates. Like in the case of the MSNCS$^{(1)}$ indicator, the length of the reference window within which active references are counted equals the length of the citation window of the publication by which the $j$th citation is received. The MSNCS$^{(2)}$ indicator is based on the idea of fractional counting of citations introduced by Leydesdorff and Opthof (2010; see also Leydesdorff & Bornmann, 2011; Leydesdorff et al., in press; Zhou & Leydesdorff, 2011).[4] However, a difference with the fractional citation counting idea of Leydesdorff and Opthof is that instead of all references in a citing publication only active references are counted.

The third source normalized indicator that we consider in our analysis is the MSNCS$^{(3)}$ indicator. In a sense, this indicator combines the ideas of the MSNCS$^{(1)}$ and MSNCS$^{(2)}$ indicators. The MSNCS$^{(3)}$ value of a journal equals

$$\text{MSNCS}^{(3)} = \frac{1}{m}\left(\frac{1}{p_1 r_1} + \frac{1}{p_2 r_2} + \ldots + \frac{1}{p_n r_n}\right), \tag{6}$$

where $r_j$ is defined in the same way as in the case of the MSNCS$^{(2)}$ indicator and $p_j$ denotes the proportion of publications with at least one active reference among all publications that appeared in the same journal and in the same year as the publication from which the $j$th citation originates. Comparing (5) and (6), it can be seen that the MSNCS$^{(3)}$ indicator is identical to the MSNCS$^{(2)}$ indicator except that $p_j$ has been added to the calculation. By including $p_j$, the MSNCS$^{(3)}$ indicator depends not only on the referencing behavior of citing publications (like the MSNCS$^{(2)}$ indicator) but also on the referencing behavior of citing journals (like the MSNCS$^{(1)}$ indicator). The rationale for including $p_j$ is that some fields have more publications without active references than others, which may distort the normalization for field differences implemented in the MSNCS$^{(2)}$ indicator. For a more extensive discussion of this issue, we refer to Waltman et al. (2012), who present a revised version of the SNIP indicator originally introduced by Moed (2010). The MSNCS$^{(3)}$ indicator is similar to this revised SNIP indicator. The main difference is that the MSNCS$^{(3)}$ indicator uses multiple citing years, while the revised SNIP indicator uses a single citing year.

---

[4] In a somewhat different context, the idea of fractional citation counting was already suggested by Small and Sweeney (1985).

## 4. Empirical analysis

Our empirical analysis is concerned with assessing the citation impact of journals in the WoS database.[5] Only journals in the sciences and the social sciences are considered. Journals in the arts and humanities are not taken into account. The period of analysis is 2008–2011. Hence, for each journal, citations received in the period 2008–2011 by publications that appeared in the period 2008–2010 are counted. Only publications of the WoS document types *article* and *review* are included in the analysis, both as cited and as citing publications.

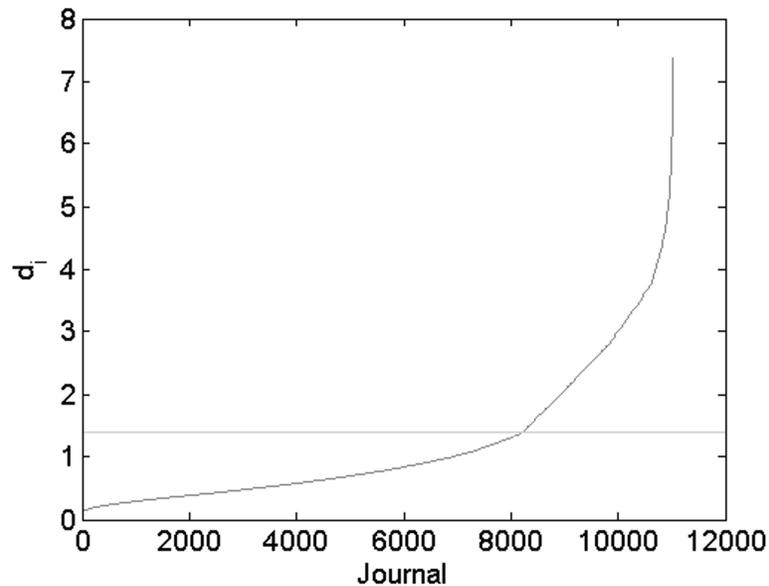

Figure 1. Distribution of journals' $d_i$ values. The horizontal line indicates the threshold of 1.3985.

The five bibliometric indicators discussed in the previous section are calculated both based on all journals in the WoS database and based on a selection of international journals. In the former case, indicator values are obtained for 11,031 journals. In the latter case, 2,816 national and regional journals are excluded from the analysis, which means that we have indicator values for 8,215 journals. The 2,816 journals are excluded because for these journals $d_i$ in (1) has a value above 1.3985, indicating that the journals have a relatively strong national or regional orientation. The threshold of 1.3985 was chosen based on three considerations. First, looking at the distribution of journals' $d_i$ values, a 'kink' was observed around $d_i = 1.4$ (see Figure 1). Second, based on a manual inspection of a sample of journals, $d_i = 1.4$ seemed a reasonable threshold for distinguishing between international journals on the one hand and national and regional journals on the other hand. And third, it was found that 1.3985 is the highest threshold for which all journals with addresses from only one country are excluded from the analysis. We note that in the case of the indicators calculated based on our selection of international journals the 2,816 national and regional journals are excluded not only on the cited side but also on the citing side. Hence, citations originating from these journals are not taken into account. We also note that non-English language publications in the 8,215 journals classified as international are excluded from the analysis as well (for a discussion of the issue of non-English language publications, see Van Raan, Van Leeuwen, & Visser, 2011a, 2011b).

---

[5] The full results of our analysis are available online at www.ludowaltman.nl/normalization/.

Table 1 lists the ten WoS subject categories with the largest number of publications in national and regional journals. Journals and publications belonging to multiple subject categories are counted fractionally in the table. Notice that the three subject categories for which the number of publications in national and regional journals is largest (i.e., 'Chemistry, Multidisciplinary', 'Medicine, General & Internal', and 'Physics, Multidisciplinary') are all special categories with a broad scope.

Table 1. Top 10 WoS subject categories with the largest number of publications in national and regional journals.

| WoS subject category | No. nat. and reg. journals | No. pub. |
|---|---|---|
| Chemistry, Multidisciplinary | 62.7 | 32,396 |
| Medicine, General & Internal | 89.7 | 31,133 |
| Physics, Multidisciplinary | 33.0 | 25,465 |
| Veterinary Sciences | 72.0 | 15,897 |
| Metallurgy & Metallurgical Engineering | 32.3 | 13,964 |
| Materials Science, Multidisciplinary | 32.0 | 11,798 |
| Physics, Applied | 12.2 | 11,506 |
| Engineering, Electrical & Electronic | 28.7 | 11,353 |
| Public, Environmental & Occupational Health | 51.8 | 10,554 |
| Pharmacology & Pharmacy | 42.6 | 10,247 |

**4.1. General statistics**

Some general statistics are reported in Tables 2, 3, and 4. Table 2 shows the Pearson and Spearman correlations for all pairs of indicators, where the indicators have been calculated based on all 11,031 journals in the WoS database. Table 3 is similar to Table 2 except that the indicator calculations are based on our selection of 8,215 international journals. We note that only journals with at least 100 publications have been included in the calculation of the correlations reported in Tables 2 and 3. Table 4 presents the average value of each of our five indicators, calculated either based on all journals or based on international journals only. In the calculation of the average values, each journal is weighted by its number of publications. For each indicator, the table also shows the Pearson and Spearman correlations between indicator values calculated based on all journals and indicator values calculated based on international journals only. These correlations are based on the indicator values obtained for international journals with at least 100 publications.

Table 2. Pearson correlations (lower left) and Spearman correlations (upper right) for all pairs of indicators. The indicators have been calculated based on all 11,031 journals in the WoS database. Only the 7,551 journals with at least 100 publications have been included in the calculation of the correlations.

|  | MCS | MNCS | MSNCS$^{(1)}$ | MSNCS$^{(2)}$ | MSNCS$^{(3)}$ |
|---|---|---|---|---|---|
| MCS |  | 0.80 | 0.79 | 0.91 | 0.75 |
| MNCS | 0.84 |  | 0.91 | 0.90 | 0.89 |
| MSNCS$^{(1)}$ | 0.87 | 0.92 |  | 0.94 | 0.98 |
| MSNCS$^{(2)}$ | 0.84 | 0.83 | 0.94 |  | 0.94 |
| MSNCS$^{(3)}$ | 0.79 | 0.85 | 0.97 | 0.98 |  |

Table 3. Pearson correlations (lower left) and Spearman correlations (upper right) for all pairs of indicators. The indicators have been calculated based on a selection of 8,215 international journals. Only the 5,820 journals with at least 100 publications have been included in the calculation of the correlations.

|  | MCS | MNCS | MSNCS$^{(1)}$ | MSNCS$^{(2)}$ | MSNCS$^{(3)}$ |
|---|---|---|---|---|---|
| MCS |  | 0.72 | 0.70 | 0.87 | 0.65 |
| MNCS | 0.82 |  | 0.88 | 0.87 | 0.86 |
| MSNCS$^{(1)}$ | 0.86 | 0.90 |  | 0.92 | 0.98 |
| MSNCS$^{(2)}$ | 0.81 | 0.81 | 0.94 |  | 0.92 |
| MSNCS$^{(3)}$ | 0.78 | 0.83 | 0.97 | 0.98 |  |

Table 4. Average value of each indicator, calculated either based on all journals or based on international journals only, and Pearson and Spearman correlations between indicator values calculated based on all journals and indicator values calculated based on international journals only. Only the 5,820 international journals with at least 100 publications have been included in the calculation of the correlations.

|  | MCS | MNCS | MSNCS$^{(1)}$ | MSNCS$^{(2)}$ | MSNCS$^{(3)}$ |
|---|---|---|---|---|---|
| Average (all journals) | 5.08 | 1.00 | 1.06 | 0.81 | 1.06 |
| Average (int. journals) | 5.43 | 1.00 | 1.06 | 0.84 | 1.06 |
| Pearson correlation | 1.00 | 0.98 | 0.99 | 1.00 | 1.00 |
| Spearman correlation | 1.00 | 0.99 | 0.98 | 0.99 | 0.98 |

Taking into account the statistics reported in Tables 2, 3, and 4, the next subsection presents a comparison of the different indicators. Subsection 4.3 considers the effect of excluding national and regional journals from the analysis.

**4.2. Comparison of indicators**

The general pattern that can be observed based on the correlations reported in Tables 2 and 3 is that the three MSNCS indicators are all quite strongly correlated, with Pearson and Spearman correlations above 0.90. The correlations of the three MSNCS indicators with the MNCS indicator are somewhat lower, but the difference is not large. The MCS indicator, which is the only indicator that makes no attempt to normalize for field differences, also has fairly high correlations with the other indicators. However, one should be careful when drawing conclusions from the correlations reported in Tables 2 and 3. The different indicators all have skewed distributions, with many journals with relatively low indicator values and only a small number of journals with high indicator values. These skewed distributions fairly easily give rise to high Pearson correlations. As we will see in the next subsection, high correlations may sometimes hide important differences between indicators.

Table 4 shows that the average MNCS value of all journals equals exactly one. This is not surprising, since this is a direct consequence of the way in which the MNCS indicator is defined. The MSNCS$^{(1)}$ and MSNCS$^{(3)}$ indicators have average values somewhat above one. In the case of these indicators, average values above one indicate that the yearly number of publications added to the database increases over time. If each year the same number of publications had been added to the database, the MSNCS$^{(1)}$ and MSNCS$^{(3)}$ indicators would have had average values very close to one (for more details, see Waltman & Van Eck, 2010b; Waltman et al., 2012). The average value of the MSNCS$^{(2)}$ indicator is substantially below one. This is a consequence of the fact that some publications have no active references. In the case of the MSNCS$^{(2)}$ indicator, publications without active references give no 'credits' to earlier publications. In this way, the balance between publications that provide credits and publications that receive credits is distorted, and this causes the average value of the MSNCS$^{(2)}$ indicator to be below one. The MSNCS$^{(2)}$ indicator would have had an average value very close to one if each year the same number of publications had been added to the database and if there had been no publications without active references (for a more extensive discussion of the issue of publications without active references, see Waltman et al., 2012).

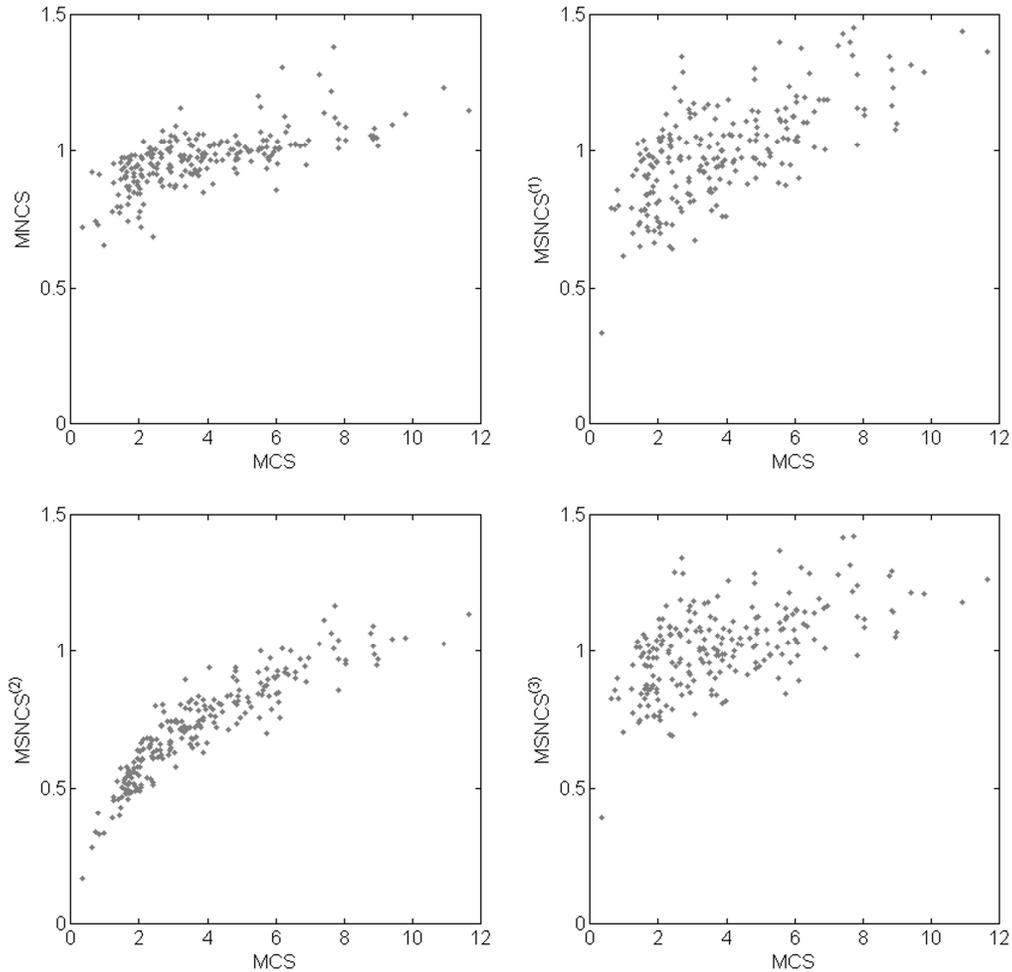

Figure 2. Scatter plots of the relations between four normalized indicators (i.e., the MNCS indicator and the three MSNCS indicators) and the unnormalized MCS indicator. The indicators have been calculated based on all journals in the WoS database. The scatter plots show average indicator values for 222 fields.

The main question of course is to what extent our indicators succeed in normalizing for field differences. This question can be answered only partially, since there is no perfectly accurate field normalization available based on which we can evaluate our indicators. To provide some insight into the normalization capabilities of our indicators, we use the field classification system provided by the WoS subject categories and we calculate for each indicator the average value of the journals in each field. In the calculation of the average values, each journal is weighted by its number of publications and journals belonging to multiple fields are treated in a fractional way. All WoS covered journals are included in the calculation, both international ones and national and regional ones. Based on the average indicator values per field, Figure 2 presents four scatter plots. Each scatter plot shows the relation at the level of fields between one of our normalized indicators (i.e., the MNCS indicator or one of the three MSNCS indicators) and the unnormalized MCS indicator. The 'multidisciplinary sciences' subject category is not included in the scatter plots, because it clearly does not represent a field and also because it has a much higher MCS value (i.e., 18.09) than the other subject categories. The scatter plots therefore include 222 fields, all from the sciences and the social sciences.

Figure 2 reveals that the $MSNCS^{(2)}$ indicator is strongly correlated with the MCS indicator. Low citation density fields, which have low MCS values, also have low $MSNCS^{(2)}$ values, while high citation density fields have high $MSNCS^{(2)}$ values. This shows that the $MSNCS^{(2)}$ indicator does not properly normalize for field differences. The problem is that low citation density fields have more publications without active references than high citation density fields. As explained above, in the case of the $MSNCS^{(2)}$ indicator, publications without active references provide no 'credits' to earlier publications. Because the proportion of these publications differs between fields, the normalization for field differences is distorted.

What may be considered remarkable in Figure 2 is that even the MNCS indicator turns out to be somewhat correlated with the MCS indicator. Given that the field classification system used in Figure 2 is the same as the one used by the MNCS indicator, one may have expected the MNCS indicator to display a perfect normalization for field differences. In that case, each field would have had an MNCS value of exactly one. The reason why the MNCS indicator does not display a perfect normalization for field differences is that WoS subject categories are partially overlapping, with some journals belonging to multiple categories. The correlation between the MNCS indicator and the MCS indicator is an artifact of this overlap.

The $MSNCS^{(1)}$ and $MSNCS^{(3)}$ indicators yield very similar scatter plots in Figure 2. This is in line with the high correlations between these two indicators reported in Table 2. Compared with the MNCS indicator, the $MSNCS^{(1)}$ and $MSNCS^{(3)}$ indicators are more strongly correlated with the MCS indicator, but the correlation is clearly weaker than in the case of the $MSNCS^{(2)}$ indicator. The correlations of the $MSNCS^{(1)}$ and $MSNCS^{(3)}$ indicators with the MCS indicator can be explained in two ways. On the one hand, the two source normalized indicators may fail to completely normalize for all field differences. This may for instance be the case if there are significant unidirectional citations flows between fields (e.g., from applied fields with a low citation density to more basic fields with a high citation density; see Waltman et al., 2012). On the other hand, however, the results shown in Figure 2 may also be due to artifacts in the WoS subject categories. If in some categories high impact journals are overrepresented while other categories have an overrepresentation of low impact journals, then the correlations visible in Figure 2 are actually to be expected.

Figure 2 makes clear that the 'fractional citation counting' approach implemented in the $MSNCS^{(2)}$ indicator does not yield satisfactory results.[6] Based on this, the $MSNCS^{(1)}$ and $MSNCS^{(3)}$ indicators seem to be preferable over the $MSNCS^{(2)}$ indicator. The choice between the $MSNCS^{(1)}$ and $MSNCS^{(3)}$ indicators appears to be of limited practical relevance, given the strong correlation between the two indicators. This is confirmed by Figure 3, which shows the relation between the two indicators at the level of journals. In the rest of this section, our focus will be mainly on the MSNCS$^{(3)}$ indicator.

---

[6] A similar conclusion is reached by Radicchi and Castellano (2012a). However, there is a fundamental difference between our analysis and the one by Radicchi and Castellano. Radicchi and Castellano apply fractional citation counting in the way it was originally proposed by Leydesdorff and Opthof (2010), which means that fractioning is done based on the total number of references in a citing publication. Instead of the total number of references, we look at the number of active references in a citing publication (cf. Leydesdorff et al., in press). Our analysis makes clear that taking into account only active references does not solve the problems of the fractional citation counting approach.

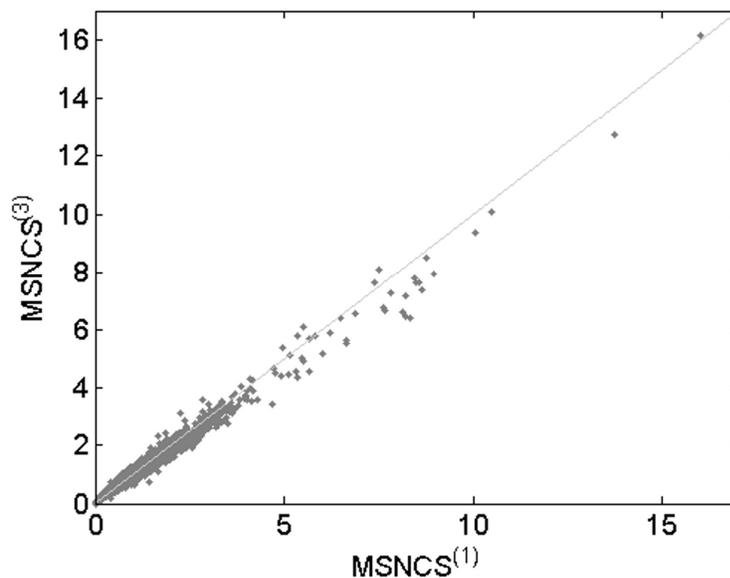

Figure 3. Scatter plot of the relation between the MSNCS$^{(1)}$ indicator and the MSNCS$^{(3)}$ indicator. The indicators have been calculated based on all journals in the WoS database. Indicator values of all journals with at least 100 publications are shown. One outlier (*Acta Crystallographica Section A*) is not visible.

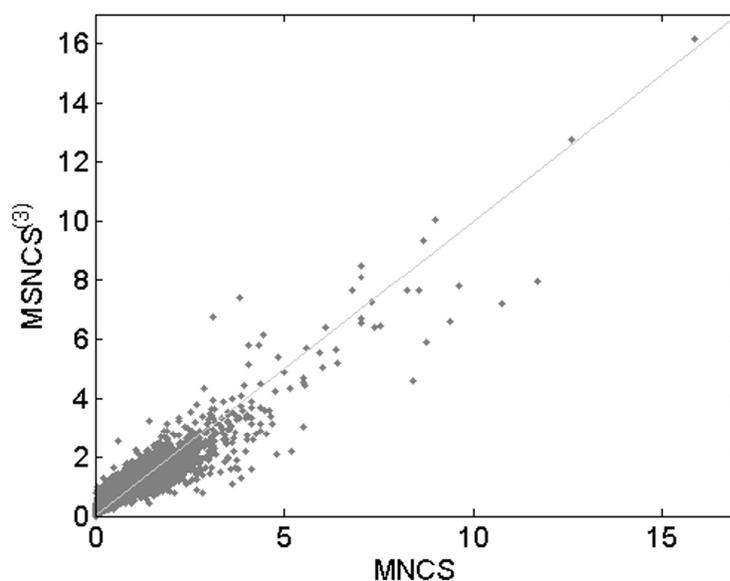

Figure 4. Scatter plot of the relation between the MNCS indicator and the MSNCS$^{(3)}$ indicator. The indicators have been calculated based on all journals in the WoS database. Indicator values of all journals with at least 100 publications are shown. One outlier (*Acta Crystallographica Section A*) is not visible.

Choosing between the MSNCS$^{(3)}$ indicator and the MNCS indicator does make a significant difference, as is shown in Figure 4 at the level of journals. Table 5 lists the journals for which the difference is largest, taking into account only journals with at least 100 publications. The left column of the table shows journals for which the MSNCS$^{(3)}$ value is much higher than the MNCS value. The right column shows journals whose MSNCS$^{(3)}$ value is much lower than their MNCS value. Although drawing general conclusions from Table 5 is difficult, some observations can be made.

Table 5. Top 15 journals with the largest positive (left column) or the largest negative (right column) difference between their MSNCS$^{(3)}$ value and their MNCS value. The indicators have been calculated based on all journals in the WoS database. Journals are listed only if they have at least 100 publications.

| Journal | Diff. | Journal | Diff. |
|---|---|---|---|
| Acta Crystallographica Section A | 28.23 | Nature Biotechnology | -3.80 |
| Science | 3.67 | Nature Materials | -3.77 |
| Nature | 3.57 | Nature Photonics | -3.57 |
| Assay and Drug Development Technologies | 1.92 | Laser Physics Letters | -3.04 |
| Cladistics | 1.80 | Nature Reviews Drug Discovery | -2.86 |
| Lancet Oncology | 1.73 | Nature Nanotechnology | -2.78 |
| Acta Crystallographica Section D | 1.67 | Psychotherapy and Psychosomatics | -2.71 |
| Clinical Microbiology Reviews | 1.45 | Journal of Informetrics | -2.59 |
| JAMA | 1.45 | Eurasian Geography and Economics | -2.58 |
| American Psychologist | 1.43 | Technological and Economic Development of Economy | -2.52 |
| Progress in Photovoltaics | 1.25 | Mass Spectrometry Reviews | -2.46 |
| PNAS | 1.17 | Veterinary Research | -2.21 |
| Journal of Turbomachinery-Transactions of the ASME | 1.14 | Pain Physician | -2.08 |
| British Medical Journal | 1.11 | International Journal of Neural Systems | -2.04 |
| Journal of the Royal Society Interface | 1.11 | Journal of Social Issues | -2.01 |

Looking at the left column of Table 5, we observe a number of journals with a broad scope. These are the multidisciplinary journals *Science*, *Nature*, *PNAS*, and *Journal of the Royal Society Interface* and the general medical journals *JAMA* and *British Medical Journal*. Because of their broad scope, journals such as these do not fit neatly into a field classification system, making it difficult for the MNCS indicator to perform a proper normalization.[7] For this reason, the MSNCS$^{(3)}$ indicator, which does not rely on a field classification system, most likely yields more accurate results for these journals.

Another special case in the left column of Table 5 is *Acta Crystallographica Section A* (MNCS = 20.88; MSNCS$^{(3)}$ = 49.11). One of the publications in this journal in 2008 has been cited extremely often.[8] In fact, more than half of all citations to publications that appeared in the WoS subject category 'Crystallography' in 2008 have been received by this particular publication. This means that on its own this publication has more than doubled the average number of citations per publication in its field, which in turn implies that in the case of the MNCS indicator this publication largely determines its own normalization. This is not the case for the $\text{MSNCS}^{(3)}$ indicator, which explains the difference between the two indicators.

---

[7] This problem is also discussed by Glänzel, Schubert, and Czerwon (1999). As a solution, these authors propose to treat journals with a broad scope in a special way. In their proposal, publications in journals with a broad scope are assigned to fields based on their references.

[8] This the following publication: Sheldrick, G.M. (2008). A short history of SHELX. *Acta Crystallographica Section A*, *64*(1), 112–122. By the end of 2011, this publication had been cited almost 25,000 times.

In the right column of Table 5, we observe *Journal of Informetrics* (MNCS = 4.15; $\text{MSNCS}^{(3)}$ = 1.56), a journal with which many readers will probably be familiar. *Journal of Informetrics* belongs to the WoS subject category 'Information Science & Library Science'. Earlier research has shown that the library and information science field is quite heterogeneous in terms of citation density (Waltman, Yan, & Van Eck, 2011). *Journal of Informetrics* is part of the subfield with the highest citation density, which causes the MNCS indicator to overestimate the journal's citation impact. It therefore seems likely that the $\text{MSNCS}^{(3)}$ indicator provides a more accurate assessment of the impact of the journal.

Based on the above observations, it can be concluded that there are at least a number of journals for which the results of the $\text{MSNCS}^{(3)}$ indicator can be expected to be more accurate than those of the MNCS indicator. A more extensive analysis is required to determine to what extent these findings generalize to other journals. We leave this as an issue for future research.

**4.3. Effect of journal selection**

We now consider the effect of excluding national and regional journals from the analysis. The Pearson and Spearman correlations reported in Table 4 are all very close to one, suggesting that there is hardly any effect. However, Figures 5 and 6 show that the high correlations may be somewhat misleading. As can be seen in Figure 6, in the case of the $\text{MSNCS}^{(3)}$ indicator, the effect of excluding national and regional journals is indeed quite small. The $\text{MSNCS}^{(3)}$ values of most journals decrease slightly. In the case of the MNCS indicator, however, Figure 5 shows a much more significant effect. For most journals, the MNCS value decreases only by a small amount, but for some journals the decrease is much larger. A number of high impact journals even lose more than half of their MNCS value. Hence, Figures 5 and 6 make clear that the MNCS indicator is considerably more sensitive to the exclusion of national and regional journals than the $\text{MSNCS}^{(3)}$ indicator. Results for the other two MSNCS indicators are not shown but are similar to those for the $\text{MSNCS}^{(3)}$ indicator.

Table 6 lists the 15 journals for which the exclusion of national and regional journals leads to the largest decrease in MNCS value. With one exception, these journals all belong to the WoS subject categories 'Chemistry, Multidisciplinary', 'Medicine, General & Internal', and 'Physics, Multidisciplinary'. As we have seen in the beginning of this section, these are the three subject categories with the largest number of publications in national and regional journals. National and regional journals usually have a relatively low citation impact. Excluding these journals therefore tends to increase the average citation impact of the publications in a field. This means that, relative to the field average, the citation impact of international journals goes down. Table 6 shows that in the case of journals in the 'Medicine, General & Internal' subject category (e.g., *New England Journal of Medicine*, *Lancet*, and *JAMA*) the decrease is even more than 50%. This clearly illustrates the sensitivity of the MNCS indicator to the selection of journals included in an analysis.

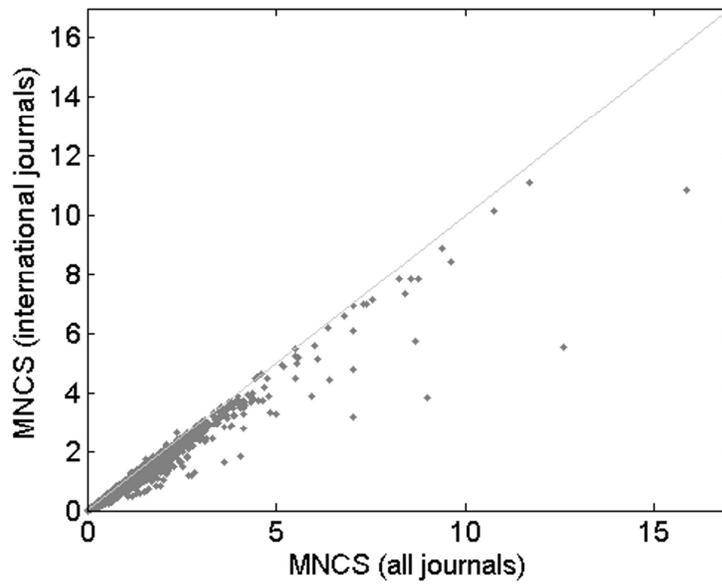

Figure 5. Scatter plot of the relation between the MNCS indicator calculated based on all WoS covered journals and the MNCS indicator calculated based on international journals only. Indicator values of all international journals with at least 100 publications are shown. One outlier (*Acta Crystallographica Section A*) is not visible.

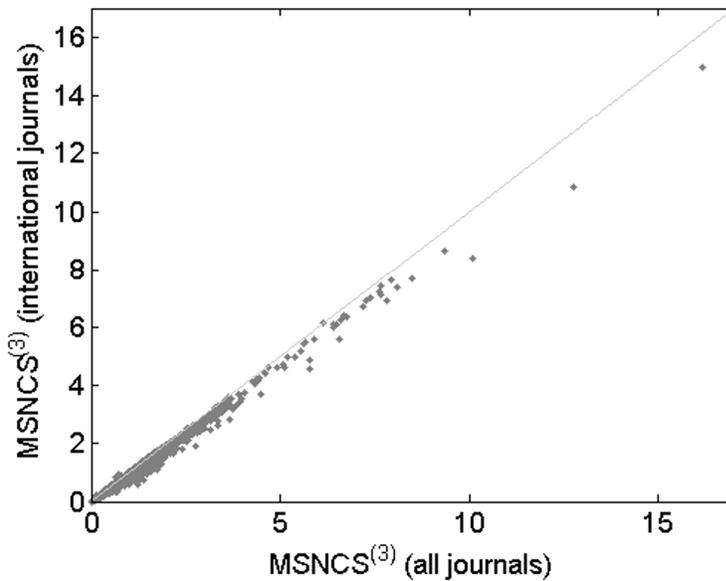

Figure 6. Scatter plot of the relation between the $MSNCS^{(3)}$ indicator calculated based on all WoS covered journals and the $MSNCS^{(3)}$ indicator calculated based on international journals only. Indicator values of all international journals with at least 100 publications are shown. One outlier (*Acta Crystallographica Section A*) is not visible.

Table 6. Top 15 journals with the largest difference between their MNCS value calculated based on all WoS covered journals and their MNCS value calculated based on international journals only. Journals are listed only if they have at least 100 publications.

| Journal | MNCS (all journals) | MNCS (int. journals) | Difference |
|---|---|---|---|
| New England Journal of Medicine | 12.61 | 5.53 | 7.08 |
| Lancet | 8.97 | 3.86 | 5.11 |
| Reviews of Modern Physics | 15.83 | 10.85 | 4.98 |
| JAMA | 7.03 | 3.16 | 3.87 |
| Chemical Reviews | 8.68 | 5.72 | 2.96 |
| Annals of Internal Medicine | 4.07 | 1.83 | 2.25 |
| Physics Reports | 7.02 | 4.80 | 2.22 |
| Chemical Society Reviews | 5.92 | 3.91 | 2.01 |
| PLoS Medicine | 3.62 | 1.62 | 2.00 |
| Nature Physics | 6.42 | 4.45 | 1.97 |
| Accounts of Chemical Research | 4.98 | 3.29 | 1.69 |
| Agricultural Systems | 2.78 | 1.19 | 1.59 |
| Archives of Internal Medicine | 2.84 | 1.28 | 1.55 |
| British Medical Journal | 2.69 | 1.16 | 1.53 |
| Reports on Progress in Physics | 4.83 | 3.31 | 1.51 |

## 5. Conclusions

We have compared a number of bibliometric indicators that differ from each other in the approach they take to normalize for differences in citation practices between scientific fields. The MNCS indicator uses a traditional normalization approach based on a field classification system, while the three MSNCS indicators that we have studied each use a different source normalization approach. We have also investigated the issue of the selection of the journals to be included in a normalization for field differences.

Based on our empirical analysis, in which we have used the different indicators to assess the citation impact of journals in the WoS database, the following conclusions can be drawn:

- The $MSNCS^{(2)}$ indicator, which is based on the idea of fractional citation counting (Leydesdorff & Bornmann, 2011; Leydesdorff & Opthof, 2010; Leydesdorff et al., in press; Zhou & Leydesdorff, 2011), does not properly normalize for field differences. Because of this, the $MSNCS^{(1)}$ and $MSNCS^{(3)}$ indicators seem to be preferable over the $MSNCS^{(2)}$ indicator.
- The $MSNCS^{(1)}$ and $MSNCS^{(3)}$ indicators, which are based on the ideas of, respectively, the audience factor (Zitt & Small, 2008) and the revised SNIP indicator (Waltman et al., 2012), are strongly correlated, and the choice between these two indicators therefore seems to be of limited practical relevance.
- The MNCS indicator has difficulties with journals with a broad scope (e.g., *Nature* and *Science*, but also *JAMA* and *British Medical Journal*) and with fields that are heterogeneous in terms of citation density (e.g., the WoS subject category 'Information Science & Library Science'). In addition, the MNCS indicator is quite sensitive to the selection of journals included in an analysis.

Overall, we think that our results provide most support to the $MSNCS^{(1)}$ and $MSNCS^{(3)}$ indicators. We acknowledge, however, that the problems observed for the MNCS indicator also relate to the field classification system that we have used (i.e., the WoS subject categories). To some extent, these problems may be solved by using a more accurate field classification system, preferably one in which fields are defined at the level of individual publications rather than at the journal level. In some disciplines, such field classification systems are available (e.g., the MeSH, PACS, and JEL systems in, respectively, biomedicine, physics and astronomy, and economics), but in many others they are not. An alternative solution therefore may be to algorithmically construct a field classification system covering all disciplines of science (e.g., Waltman & Van Eck, in press).

There are a number of important issues for future research. In particular, we would like to mention the following topics:

- There is a clear need for additional empirical work in which comparisons are made between normalization approaches based on field classification systems and source normalization approaches. These comparisons could for instance zoom in on individual scientific fields. Also, instead of journals, they could focus on other units of analysis, such as individual researchers, research groups, or universities. The most rigorous approach to comparing normalization approaches would be to investigate the extent to which different approaches produce universal patterns in the normalized citation distributions of scientific fields.
- Criteria need to be developed for distinguishing between different types of journals, such as 'regular' scientific journals, scientific journals with a strong national or regional focus, trade journals, and popular magazines. Using such criteria, certain types of journals can be excluded from a normalization for field differences.[9] When taking a source normalization approach, it is especially important to exclude journals with very small numbers of active references (Waltman et al., 2012).
- As already suggested above, the MNCS indicator needs to be tested with other field classification systems. Ideally, a field classification system would be used that is defined at the level of individual publications and that covers all disciplines of science.
- The effect of different normalization approaches on different families of indicators needs to be investigated. In this paper, our focus has been exclusively on average-based indicators. An alternative possibility could be to investigate indicators that are based on the idea of counting highly cited publications.

## Acknowledgment

We are grateful to Javier Ruiz Castillo for his comments on an earlier draft of this paper.

---

[9] Recent studies on classification-system-based normalization approaches focus on identifying general patterns in the citation distributions of scientific fields (e.g., Crespo, Li, & Ruiz-Castillo, 2012; Radicchi & Castellano, 2012b; Radicchi, Fortunato, & Castellano, 2008). These studies usually do not exclude any journals. It seems likely that the results of these studies depend quite significantly on whether trade journals, popular magazines, and other special journals are included or excluded.